\documentclass[aps,prl,twocolumn,groupedaddressm,floatfix]{revtex4}
\usepackage{graphicx}

\begin{document}

\title{Controlling the coexistence of structural phases and the optical properties of gallium nanoparticles with optical excitation}

\author{K. F. MacDonald}
\affiliation{School of Physics and Astronomy, University of
Southampton, SO17 1BJ, UK}
\author{V. A. Fedotov}
\affiliation{School of Physics and Astronomy, University of
Southampton, SO17 1BJ, UK}
\author{S. Pochon}
\affiliation{School of Physics and Astronomy, University of
Southampton, SO17 1BJ, UK}
\author{G. Stevens}
\affiliation{School of Physics and Astronomy, University of
Southampton, SO17 1BJ, UK}
\author{F. V. Kusmartsev}
\affiliation{Department of Physics, Loughborough University, LE11
3TU, UK}
\author{N. I. Zheludev}
\affiliation{School of Physics and Astronomy, University of
Southampton, SO17 1BJ, UK}



\begin{abstract}
 We have observed reversible structural transformations, induced by
optical excitation at 1.55 $\mu m$, between the $\beta$, $\gamma$
and liquid phases of gallium in self-assembled gallium
nanoparticles, with a narrow size distribution around 50 nm, on
the tip of an optical fiber. Only a few tens of nanowatts of
optical excitation per particle are required to control the
transformations, which take the form of a dynamic phase
coexistence and are accompanied by substantial changes in the
optical properties of the nanoparticle film. The time needed to
achieve phase equilibrium is in the microsecond range, and
increases critically at the transition temperature.
\end{abstract}

\maketitle

We report that the coexistence of different crystalline and
disordered phases in nanoparticles can be controlled by optical
excitation in a \emph{controllable}, \emph{continuous} and
\emph{reversible} fashion. This has been observed in nanoparticles
of polymorphic elemental gallium. Shifts in the balance between
phases are accompanied substantial changes in the optical
properties of the nanoparticle film. The light-induced
transformations display phenomenological features typical of
1st-order Ehrenfest phase transitions (such as a reflectivity
hysteresis) but simultaneously display characteristics of
2nd-order transitions (such as critical dependencies on
temperature of the `susceptibility' and relaxation time of the
stimulated response). Our study is motivated by a desire to
understand the exciting physics of phase equilibria in
nanoparticles \cite{Smirnov94-PS50, Berry84-PRA30, Berry00-JCP113}
and in particular in metallic nanoparticles, which have the
potential to play a key role in future highly integrated photonic
devices as the active elements of waveguiding \cite{Krenn99-PRL82}
and switching \cite{Zheludev02-CP43} structures.

We studied light-induced structural transformations in gallium
nanoparticles by monitoring the optical reflectivity of
nanoparticle films. The particles, typically 50 nm in diameter
with a relatively narrow size distribution ($\pm14$ $nm$), were
prepared on the tips of silica optical fibers, using the recently
developed light-assisted self-assembly technique
\cite{MacDonald02-APL80,Fedotov03-JAP93}. This process produced a
nanoparticle film on the fiber's core ($9$ $\mu m$ in diameter)
comprising $\sim$$2.10^4$ nanoparticles. Phase transitions in the
nanoparticles ware stimulated by a $1.4$ $mW$, $1.55$ $\mu m$
diode laser launched into the fiber. Its output was modulated with
$50\%$ duty cycle at $6.2$ $kHz$ providing an average excitation
power of about 22 $nW$ per nanoparticle. Another $0.4$ $mW$ cw
diode laser operating at $1.31$ $\mu m$ and phase-sensitive
detection apparatus were used to monitor the reflectivity of the
film. The detection system had an overall bandwidth of $300$
$kHz$.

\begin{figure}[!h]
    \includegraphics[width=85mm]{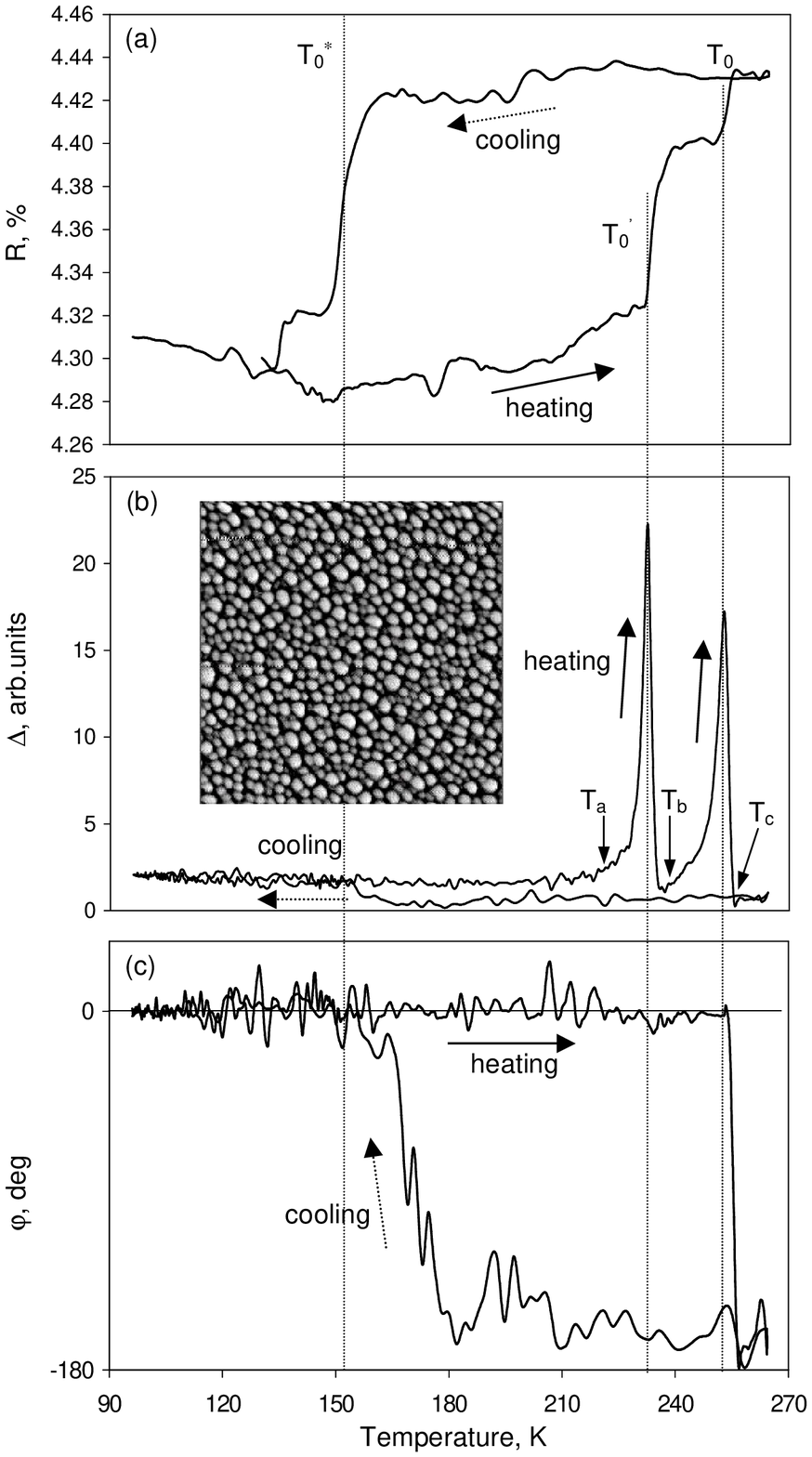}\\
    \caption{Temperature dependencies of (a) Reflectivity, (b) pump-induced reflectivity change
$\Delta$, and (c) phase-shift $\phi$ between pump excitation and
probe response for gallium nanoparticles on the core of a single
mode optical fiber. The inset to (b) shows an atomic force
microscope image of a $2 \times 2$ $\mu m$ area of
nanoparticles.}\label{Fig1}
\end{figure}

The nanoparticle film's reflectivity shows a very wide ($>$$100$
$K$) hysteresis, with two distinct steps, at $T_0^\prime \sim 233$
$K$ and $T_0 \sim 253$ $K$, in the rising temperature part of the
cycle (Fig.~\ref{Fig1}a). The presence of these steps indicates
that there are two structural phase transitions and therefore that
either the nanoparticles undergo a transition from one solid phase
to another and then from that phase to the liquid, or that two
solid phases coexist at low temperatures and undergo the
transition to liquid at different temperatures. Analysis of
gallium's phase diagram \cite{Bosio78-JCP68} and results from
energy-dispersive x-ray diffraction studies of gallium
nanoparticles \cite{DiCicco98-PRL81} suggest that in the present
case the solid phase with the highest melting point is
$\beta$-gallium, and that the other solid phase is
$\gamma$-gallium.

\begin{figure}[!h]
    \includegraphics*[width=85mm]{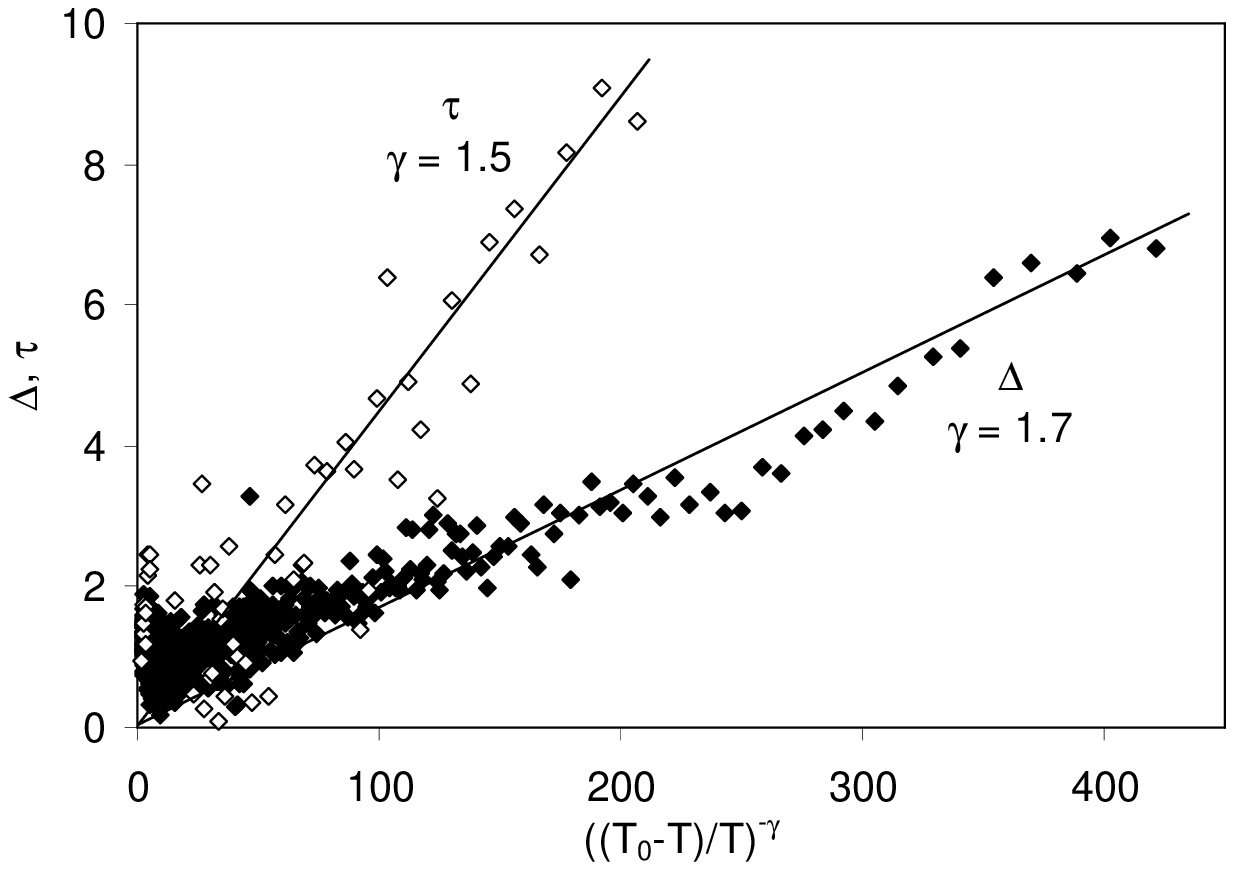}
    \caption{Pump-induced
reflectivity change $\Delta$ and relaxation time $\tau$ as scaling
functions of temperature.}\label{Fig2}
\end{figure}

The presence of pump excitation at $1.55$ $\mu m$ changes the
reflectivity of the nanoparticle film by several percent in a
reversible and reproducible fashion. The magnitude $\Delta$ of the
induced reflectivity change (Fig.~\ref{Fig1}b) has a non-zero
value at all temperatures, with pronounced peaks that coincide
with the steepest parts of the corresponding reflectivity versus
temperature plot. With increasing temperature, and the relative
phase shift (Fig.~\ref{Fig1}c) between the pump modulation and the
change in probe reflectivity $\phi \approx$ $0$ across the whole
range from $90$ $K$ to $\sim$$253$ $K$ $(T_0)$, indicating that
the reflectivity responds quickly to optical stimulation, with
relaxation times smaller than $10$ $\mu s$. At $T_0$ and
$T_0^\prime$, $\Delta$ increases by a factor of $\sim$$8$, and
above $T_0$ the phase shift $\phi$ becomes negative. Above
$\sim$$255$ $K$, $\phi \approx -180^\circ$, meaning that optical
excitation actually reduces the reflectivity of the nanoparticles
in the molten phase. The relaxation time $\tau$ of the
nonlinearity increases steeply in the proximity of the peak with
scaling dependence $\tau \propto (T_0-T)^{-1.7}$, while the
magnitude of the response in the same temperature range follows
$\Delta \propto (T_0-T)^{-1.5}$ (Fig.~\ref{Fig2}). Remarkably, the
phase shift between the pump modulation and the probe reflectivity
change remains almost zero at $T_0^\prime$, indicating that the
response and relaxation times remain short ($<10$ $\mu s$). With
decreasing temperature, the pump and probe remain in anti-phase
until $\sim$$180$ $K$, then $\phi$ returns to zero by $\sim$$150$
$K$. A small, smooth peak is seen in $\Delta$ at the
re-crystallization temperature $T_0^*$.

\begin{figure}[!h]
    \includegraphics{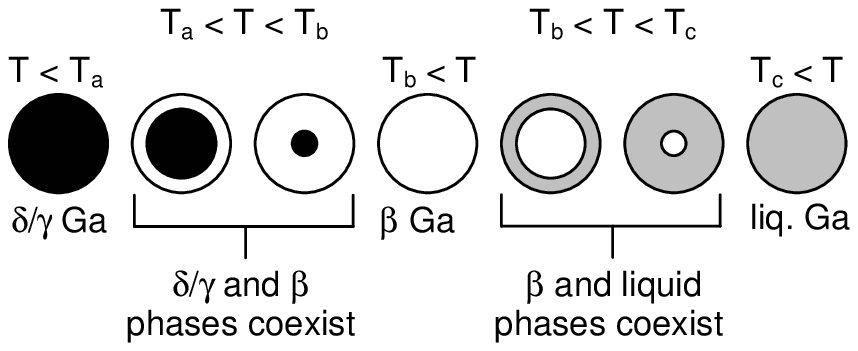}
\caption{Shell model for a solid-solid-liquid series of structural
transformations occurring with increasing temperature in gallium
nanoparticles.}\label{Fig3}
\end{figure}

The phenomenological picture seen here is unusual and is quite
different from what is normally observed in the bulk where first
order phase transitions are characterized by a discontinuous
change in the state of the body, effected by means of a sudden
rearrangement of the crystalline lattice. Many features of our
experimental results may be explained by consideration of the fact
that under the confined conditions of the nanoparticle geometry,
there is a continuous dynamic coexistence of structural forms in
the nanoparticles. Such behavior has been studied theoretically
for small clusters, and it was predicted that there would be a
temperature interval wherein two different phases coexist
\cite{Berry84-PRA30,Smirnov94-PS50}. The surface of a particle,
where atoms have fewer nearest neighbors than internal atoms,
becomes very important and acts as a boundary at which
transformation processes can start. We believe that this concept
is largely applicable to the present case and it brings us to a
model in which we consider an ellipsoidal nanoparticle with a core
in one structural phase covered by a homogeneous `shell' of
another phase. In the simplest case of a phase transition to the
melt, the light-induced behavior of such a particle would be
analogous to the temperature-driven `surface melting' effect that
has been seen in lead nanoparticles \cite{Garrigos89-ZPD12} and
found to be thermodynamically reversible within a narrow
temperature range \cite{Peters97-APL71}. In the presence of light,
the equilibrium position of the phase boundary, or to put it
another way, the thickness of the surface layer, will be
determined by both temperature and light intensity. At a certain
temperature $T_a$ below the phase transition point $T_0^{'}$ the
influence of light on the surface layer's thickness becomes
apparent as optical excitation changes the reflectivity of the
film (see Fig.~\ref{Fig1}b). With increasing temperature or level
of optical excitation, the surface layer's thickness increases
until the transformation of the core to the `surface' phase is
completed. When, at $T_b$, the core of the particle is fully
consumed by the new phase the nanoparticle becomes stable against
a return to the old phase because this would require the creation
of a nucleation center. However, if the temperature or level of
optical excitation is reduced \emph{before} the transformation to
the new phase is complete, i.e. while a nucleus of the old core
phase is still present, the transformation is reversed and the
skin layer shrinks to an appropriate equilibrium position. Thus,
reversibility is provided in the temperature range between $T_a$
and $T_b$. This whole process is then replicated around the next
phase transition temperature $T_0$ (see Fig.~\ref{Fig3}a). From
the standpoint of phase transition theory the observed effect
shows some features of second-order transitions: the thickness of
the surface layer is characterized by the correlation length of
near surface fluctuations $\xi \propto (T_0-T)^{-\nu}$. The
susceptibility is then a scaling function of the correlation
length: $\chi = (\delta \eta / \delta h)_T \propto
(T_0-T)^{-\gamma_1}$, where $\gamma_1 =1.5$ in our experiment.
When the optical excitation is terminated, the skin layer retreats
and the reflectivity is restored to its original level. Like the
susceptibility, the relaxation time has scale-invariant form, i.e.
$\tau \propto (T_0-T)^{-\gamma_3}$, where $\gamma_3 =1.7$ in our
experiment. The longer relaxation times near the phase transition
temperatures result from the increased thickness of the surface
layer and reduced velocity of the interface between the phases. In
terms of Landau theory this situation may be described by taking
the free energy of a particle to be a polynomial function of the
order parameter $\psi \sim r^{1/3}$, where $r$ is the radius of
the solid core within a particle of radius $R$. As is usual in the
vicinity of a second order phase transition, the value of the
order parameter will characteristically be proportional to $(T_0 -
T)^{1/2}$, giving $r \propto (T_0 - T)^{1.5}$. The reflectivity of
the nanoparticle film will then, to a first approximation, be
proportional to the volume of the molten phase: $V_m = 4 \pi R^3
[(R-r)/R-((R-r)/R)^2+{((R-r)/R)^3}/3] \sim t^{1.5}$, where
$t=-(T-T_0)/(T_0)$, thus producing correct estimates for the
experimental values of the critical indices. Intriguingly, the
relaxation time for the first transition at $T_0^{'}$ is much
shorter than for the second one at $T_0$. This may be an
indication that the first transition occurs between two solid
phases and would be consistent with the thermodynamic argument
that the enthalpic and entropic contributions to the Gibbs free
energy appear to enter differently into the kinetics of
solid-solid and solid-liquid transitions because of the difference
in the configurational and vibrational space available in the
solid and liquid phases \cite{Tsao86-PRL56}.

The strength of the phase coexistence concept is supported by
calculations of the optical properties of gallium nanoparticle
films on a dielectric substrate performed using a recently
developed effective-medium model for densely packaged nanoshells
\cite{Fedotov03-JOPTA6}. For the purposes of our calculations, the
dielectric constants of $\beta-$ and $\gamma-$ gallium, which are
much closer to those of a free-electron metal than those of the
$\alpha$ phase, were estimated by using the damping constant in
Drude's free-electron model as a fitting parameter to produce the
nanoparticle film reflectivity levels shown in Fig.~\ref{Fig2}a.
These calculations (also detailed in Ref.
\cite{Fedotov03-JOPTA6})confirmed that the presence on each
nanoparticle of a shell just a few nanometres thick in a phase
different from the core can produce a change in reflectivity
sufficient to explain our experimental data.

A thermally activated transition due to laser-induced heating can
explain certain characteristics of the effect. For instance, by
assuming a local light-induced temperature increase of $0.42$ $K$,
one can derive a good facsimile of the experimental peaks in
$\Delta$ at $T_0^{'}$ and $T_0$ from the reflectivity data in
Fig.~\ref{Fig1}a. However, there are serious discrepancies between
the results of this thermal model and the experimental results,
primarily at temperatures more than a few degrees below the peaks,
where the observed effect is larger than predicted. The thermal
model also fails to explain important details of $\phi$'s
dependence on temperature. This suggests that another,
temperature-independent, non-thermal excitation mechanism is also
contributing to the effect. This mechanism may be especially
important for gallium which is know to have covalent bonds within
some of its crystalline structures \cite{Bernasconi95-PRB52}. The
covalent bonds' absorption line encompasses the pump wavelength
\cite{Gong91-PRB43}, so optical excitation may result in
bonding-antibonding transitions, which destabilize the structure
\cite{Siegal95-ARMS25}. A `defect' or `inclusion' of a new phase
is thus created, changing the optical properties of the `host'
phase at temperatures far below its transition point. Furthermore,
the latent heat of transition from the defect-containing structure
to a new phase is reduced in proportion to the number of defects.
The latent heat controls the balance between phases
\cite{Trittibach94-PRB50}, so this defect creation eventually
shifts the phase equilibrium and promotes the formation of a
thicker layer of the new phase without any increase in
temperature. This process could provide an additional, non-thermal
contribution the peak at $T_0$ and $T_0^{'}$. Such `optical
melting' mechanisms, proceeding through light-induced
destabilization of the crystalline bond structure, have been
analysed previously for gallium \cite{MacDonald01-JOSAB18} and
selenium \cite{Poborchii99-APL74}.

In conclusion, we have found that light can stimulate reversible
structural transitions in gallium nanoparticles that belong to a
novel class of surface-driven excitation-induced phase transitions
and have some striking similarities with a transition recently
observed in superfluid helium droplets \cite{Kimball02-ConfLTP}
and a transformation believed to occur in confined
superconductors, wherein the existence of a `boundary phase',
characterized by spontaneous fluctuational nucleation of
short-lived vortices is critical \cite{Sobnack01-PRL86}. If there
are substantial differences between the optical properties of the
phases involved, as there are in gallium, this type of transition
provides a new means of obtaining a large optical nonlinearity,
which may be used to control light with light in nanostructured
materials such as `plasmon waveguides' made from arrays of
metallic nanoparticles.

The authors acknowledge the financial support of the EPSRC (UK).

\end{document}